# Wurtzite III-nitride distributed Bragg reflectors on Si (100) substrates

M. A. Mastro, R. T. Holm, N. D. Bassim, C. R. Eddy, Jr., R. L. Henry, M. E. Twigg
U.S. Naval Research Laboratory, Electronics Science and Technology Division; 4555 Overlook Ave., SW, Washington, D.C. 20375, USA

A. Rosenberg
U.S. Naval Research Laboratory, Optical Sciences Division; 4555 Overlook Ave., SW, Washington, D.C. 20375

**Abstract**

Distributed Bragg reflectors (DBRs) composed of an AlN/GaN superlattice were demonstrated for the first time on Si (100) substrates. Single-crystal wurtzite superlattice structures were achieved on this cubic substrate by employing offcut Si (100) wafers with the surface normal pointing 4° towards the [110] direction. This misorientation introduced an additional epitaxial constraint that prevented the growth of a two-domain GaN surface as well as cubic GaN inclusions. A crack-free 600 nm GaN cap / 5x AlN / GaN DBR structure on Si (100) was demonstrated. This accomplishment of a wurtzite III-nitride DBRs on Si (100) opens the possibility to integrate novel optical and optoelectronic devices with established Si microelectronics technology.

Monolithic integration of compound semiconductor films with Si microelectronics has been perpetually hampered by the inherent dissimilarity in crystal structures. Recently, there has been encouraging success in incorporating selective epitaxy of SiGe in the source/drain regions to create longitudinal uniaxial compressive strain to enhance the mobility of p-type MOSEFTs,[1] as well as recent developments in replacing the channel region with more exotic compounds such as InSb to create ultra-high mobility quantum well transistors that operate at low power.[2] Comparatively, the development of compound semiconductor optical or optoelectronic devices for insertion into Si microelectronics has been hindered by the greater complexity in such structures. Still, there would be a huge performance advantage in combining emitters, detectors, and waveguides as well as, potentially, high power electronics onto a Si integrated circuit versus the current practice of connecting discrete components from external sources, i.e., off-chip.[3]

Unfortunately, Si and SiGe are indirect semiconductors and thus it is difficult to fabricate efficient optical transmitters that are self-contained, i.e., without the necessity of an external source.[4] Alternatively, InGaAsP has a direct bandgap over a large composition range which allows the fabrication of efficient laser diodes and sensitive photodetectors; however, these devices are critically sensitive to structural defects that are unavoidably generated during growth on mismatched substrates such as Si. An attractive alternative is the III-nitride semiconductor system that can be tailored to emit, absorb or transmit light from the ultraviolet to the infrared, possesses a high electro-optic coefficient, and, for optoelectronics with InGaN active regions, is less sensitive to structural defects than traditional semiconductor devices.[5,6]

Nevertheless, epitaxy of GaN on Si (100) is difficult due to the large difference in thermal expansion coefficient and lattice mismatch with the substrate and, particularly, the cubic nature of Si compared to the more thermodynamically stable wurtzite (hexagonal) structure of III-nitrides. The Si (100) crystal presents a 4 – fold symmetry that is unsuitable for deposition of wurtzite GaN with a 6-fold symmetry. Deposition of GaN on on-axis Si (100) leads to the formation of two types of wurtzite GaN domains rotated by 90° as well as inclusions of cubic GaN. Recent work has shown that employing a Si (100) substrate that is offcut by 4 to 5 degrees will introduce an asymmetric constraint that allows for deposition of single-crystal wurtzite GaN.[7,8]

Deposition of GaN on Si is plagued by a large difference in thermal expansion between the film and the substrate that generates a high density of cracks when the structure is cooled from growth to room temperature. On Si (111), several groups employed a superlattice (SL) or graded AlGaN layer inserted near the AlN buffer to introduce a compressive stress during growth that compensates for the large tensile thermal stress.[9,10] The large lattice mismatch between GaN and Si is known to generate a high level of dislocations ($>10^{12}$ cm$^{-2}$) during the initial stages of growth.[11,12] The use of a offcut Si (100) substrate further exacerbates this issue by introducing a large number of step-edges which disturbs the lattice ordering during the initial stages of III-nitride growth.[13] Despite these impediments, III-nitride light emitting diodes and high electron mobility transistors have been recently demonstrated on Si (100) substrates;[7,8] however, there have been no reports of distributed Bragg reflectors (DBR) on Si (100). The ability to create III-nitride DBRs on Si (100) would open the possibility for improved design of devices for the generation, detection and propagation of ultraviolet and visible light including laser diodes, detectors and waveguides.

This letter reports on metalorganic chemical vapor deposition (MOCVD) of an AlN/GaN SL directly on Si (100) substrates offcut by 4° towards the [110] direction. The alternating sequence functions optically as a high-reflectance DBR and structurally as a strain compensating SL. Growth was carried out in a modified vertical impinging flow chemical vapor deposition reactor. Two-inch Si wafers were cleaned via a modified Radio Corporation of America process followed by an *in situ* H$_2$ bake. An Al seed layer was deposited prior to the onset of NH$_3$ flow to protect the Si surface from nitridation. The AlN/GaN SL was deposited at 1050 °C and 50 Torr. *In situ* spectroscopic interferometry monitored the layer thickness and reflectance in real time. The stop band for the DBR structure can be tuned to any point in the near-ultraviolet or visible by adjusted the thicknesses of the AlN and GaN layers in the DBR. For example, a DBR with a stop band centered at 495 nm (blue-green) would require alternating 60-nm AlN and 51-nm GaN layers. The GaN cap layer was deposited at 1020 °C at 250 Torr directly on the SL. The higher deposition pressure resulted in GaN layers with larger grains and lower defect densities. Details of this procedure and apparatus have been reported elsewhere.[14]

Structural characterization was performed with a Hitachi H-9000 top-entry transmission electron microscope operated at 300 kV and a Panalytical X'pert x-ray diffraction (XRD) system. The *ex situ* reflectance was measured at normal incidence using a halogen lamp as a source and reflected beam was dispersed through an Ocean Optics S2000 spectrometer with a 50 $\mu$m slit.

In this work, sharp interfaces were found in the SL as displayed in Fig. 1, which is an electron micrograph of a 600 nm GaN cap / 5× AlN/GaN DBR on Si (100) structure. Additionally, the alternating sequence of AlN and GaN in a SL acts to filter dislocations originating from the interface. Transmission electron microscopy (not shown) found an extremely high level of threading dislocations at the Si/AlN interface ($>10^{13}$ dislocations/cm$^2$). The dislocation level was observed to drop by more than two orders of magnitude through the SL. A similar dislocation annihilation mechanism was reported by the authors for DBR structures grown on Si (111).[11] Still, the dislocation density in the GaN cap layer on Si (100) is approximately an order of magnitude higher than that reported for a GaN cap layer on Si (111).[12] A more detailed comparison is in progress and will be reported elsewhere.

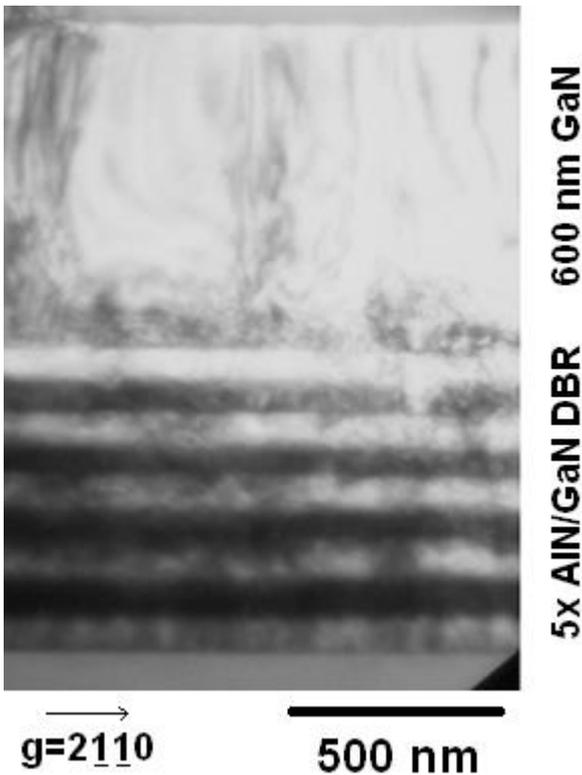

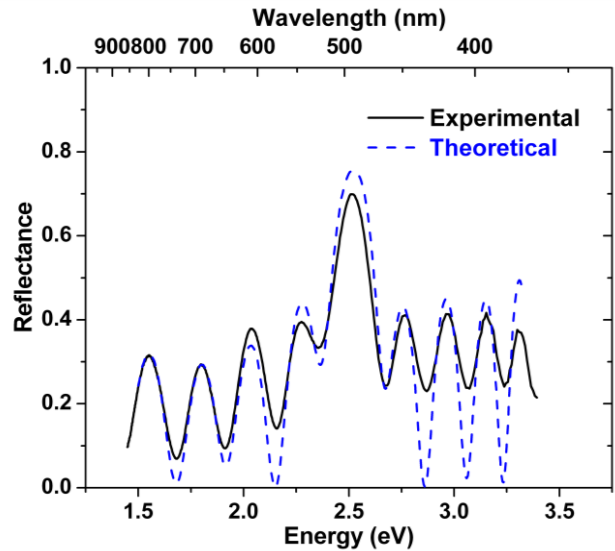

Fig. 2. (Color Online) Optical reflectance of a 600-nm GaN cap / 5× AlN/GaN DBR on Si (100) displaying 69.8% experimental and 75.1% theoretical reflectance for a stop band centered at 495-nm

Fig. 1. Dark-field transmission electron micrograph of nominally crack-free 600-nm GaN cap / 5x AlN/GaN DBR structure on a Si (100) substrate. Sharp interfaces are evident in SL despite the high level of dislocations in the entire structure.

The theoretical and experimental reflectance of a 600 nm GaN cap / 5x DBR on Si(100) is displayed in Fig. 2. The theoretical reflectance was calculated via the standard transfer matrix method[15] for a stop band centered in the blue-green portion of the visible spectrum at 495 nm. The wavelength (or energy) dependent refractive indices were reported Mastro et al.[11] To estimate the reflectance, the calculation assumed an ideal SL for the particular stop band. An experimental primary reflectance of 69.8% at 495-nm is compared against a theoretical primary reflectance of 75.1% in Fig. 2. For the DBR on Si (100), the relatively lower reflectance of the experimental reflectance compared to the theoretical reflectance is attributed to scattering due to roughness at the surface and absorption due to defects in the material as well as alloy mixing within the DBR layers.

Strong wurtzite GaN (0002) and (0004) peaks in the $\omega$-$2\theta$ x-ray diffraction (XRD) scan are observable in Fig. 3a. The diffraction intensity from the GaN cap layer is much larger than the diffraction intensity from GaN layers in the 5x DBR. Thus, the intensity of GaN cap diffraction obscures the diffraction signal of the GaN layers in the DBR. An expansion along $\omega$ revealed a full width at half maximum (FWHM) of 0.92° for the GaN (0004) peak. The wide FWHM of the GaN (0004) $\omega$ rocking curve is attributed to the high level of lattice distortion and dislocations in the GaN cap layer. A broadening of the $\omega$ (0004) rocking curve, as was seen in this film, is typically attributed to slight distortions (tilt) between grains in the film.[6] Additionally, the AlN (0004) diffraction peak is observable but is not clearly defined. The weakness of the AlN (0004) diffraction is attributed to poor material quality while the broadening of the reflection is attributed to alloy mixing that occurred during AlN and GaN layer growth in the DBR.

In summary, a novel MOCVD grown III-nitride DBR structure on Si (100) was demonstrated. The reflectance of the III-nitride DBR was enhanced by growing the superlattice directly on a Si substrate to augment the overall reflectance due to the high index of refraction contrast at the Si/AlN interface. The DBR structure consisted of alternating layers of AlN and GaN that introduced a compressive stress to balance the large tensile stress generated during cool down from growth temperature. This report demonstrates that III-nitride DBR structures can be fabricated on Si (100) substrates and conceivably within Si integrated circuits.

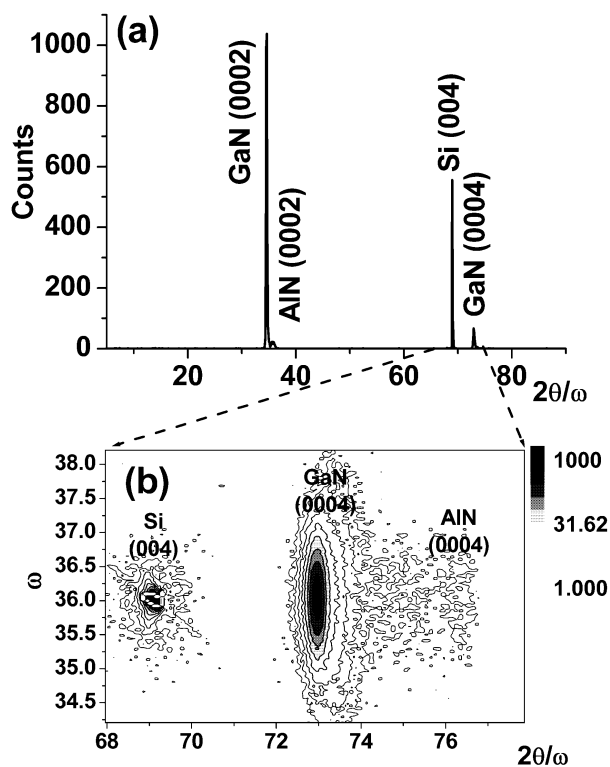

Fig. 3. X-ray diffraction of a single-crystal 600 nm GaN / 5x DBR / Si (100) substrate. (a) Strong wurtzite GaN (0002) and (0004) peaks are observable at approximately 34.4° and 73.2° in the ω-2θ scan. No other phases of GaN were detected including cubic GaN within the sensitivity of X-ray diffraction. (b) Expansion of ω-2θ scan along ω reveals a well-defined Si (400) peak from the substrate and a broad GaN (0004) peak predominantly from the 600 nm GaN cap layer. The weak and disperse AlN peak is attributed to poor material quality and alloy mixing in the AlN layers.

**Acknowledgements**

Research at the Naval Research Lab is supported by Office of Naval Research; support for two of the authors (M.A.M. and N.D.B.) was partially provided by the American Society for Engineering Education. The authors thank Mohammad Fatemi for technical discussions.

**References**
1. S.E. Thompson, M. Armstrong, C. Auth, S. Cea, R. Chau, G. Glass, T. Hoffman, J. Klaus, Z. Ma, B. Mcintyre, A. Murthy, B. Obradovic, L. Shifren, S. Sivakumar, S. Tyagi, T. Ghani, K. Mistry, M. Bohr, and Y. El-Mansy, IEEE Electron Device Letters, **25**, 191 (2004).
2. T. Ashley, A.R. Barnes, L. Buckle, S. Datta, A.B. Dean, M.T. Emery, M. Fearn, D.G. Hayes, K.P. Hilton, R. Jefferies, T. Martin, K.J. Nash, T.J. Phillips, W.A. Tang, P.J. Wilding, and R. Chau, *Proceedings on the 7th International Conference on Solid-State and Integrated Circuits Technology*, **3**, 2253 (2004).
3. S. I. Assoc. International Technology Roadmap for Semiconductors. Technical report, 2001.
4. H. Rong, R. Jones, A. Liu, O. Cohen, D. Hak, A. Fang, and M. Paniccia, Nature **433**, 725 (2005).
5. M. Henini and M. Razeghi, *Optoelectronic Devices: III Nitrides* (Elsevier Science, Berlin, 2005).
6. M.A. Mastro, D.V. Tsvetkov, A.I. Pechnikov, V.A. Soukhoveev, G.H. Gainer, A. Usikov, V. Dmitriev, B.Luo, F. Ren, K.H. Baik, and S.J. Pearton, Mat. Res. Soc. Proc. **764**, C2.2 (2003).
7. S. Joblet, F. Semond, F. Natali, P. Vennegues, M. Laugt, Y. Cordier, and J. Massies, Phys. Stat. Sol. (c), **2-7**, 2187 (2005).
8. A. Mills, III-Vs Review, **19-1**, 25 (2006).
9. M.A. Mastro, C.R. Eddy Jr., D.K. Gaskill, N.D. Bassim, J. Casey, A. Rosenberg, R.T. Holm, R.L. Henry, and M.E. Twigg, J. Crystal Growth, **287**, 610 (2006)
10. A. Reiher, J. Bläsing, A. Dadgar, A. Diez, and A. Krost, J. Cryst. Growth **248**, 563 (2003).
11. M.A. Mastro, R.T. Holm, N.D. Bassim, C.R. Eddy Jr., D.K. Gaskill, R.L. Henry, M.E. Twigg, Appl. Phys. Lett., **87**, 241103 (2005).

12. M.A. Mastro, R.T. Holm, N.D. Bassim, D.K. Gaskill, J.C. Culbertson, M. Fatemi, C.R. Eddy Jr., R.L. Henry, and M.E. Twigg, J. Vac. Sci. and Tech. B (to be published).
13. N. D. Bassim, M. E. Twigg, C. R. Eddy Jr., J. C. Culbertson, M. A. Mastro, R. L. Henry, R. T. Holm, P. G. Neudeck, A. J. Trunek, and J. A. Powell, Appl. Phys. Lett. **86**, 21902 (2005).
14. C. R. Eddy, Jr., R. T. Holm, R. L. Henry, J. C. Culbertson, and M. E. Twigg, J. Electron. Mater. **34**, 1187 (2005).
15. M. Born and E. Wolf, *Principles of Optics* (Pergamon, New York, 1959).